# Raman Excited Spin Coherences in N-V Diamond

**P.R. Hemmer,**

Air Force Research Laboratory, Sensors Directorate, Hanscom AFB, MA 01731

**A.V. Turukhin and M. S. Shahriar**

Research Laboratory of Electronics, Massachusetts Institute of Technology,

77 Massachusetts Avenue, Cambridge, MA 02139

**J. A. Musser**

Texas A&M University

College Station, TX 77843

**Abstract:** Raman excited spin coherences were experimentally observed in N-V diamond color centers via nondegenerate four-wave mixing (NDFWM) and electromagnetically induced transparency (EIT). The maximal absorption suppression was found to be 17%, which corresponds to 70% of what is possible given the random geometric orientation of the N-V center in diamond. In the context of quantum computing in solids, this level of transparency represents the efficient preparation of quantum bits (qubits), as well as ability to perform arbitrary single qubit rotations.

The use of optical Raman interactions to excite spin coherences in solid materials has numerous potential applications, ranging from low-power nonlinear optics[1] to high-temperature spectral hole burning memories[2] to solid-state quantum computing[3]. The interest in Raman excitation lies in the fact that the spin coherences can be efficiently excited and manipulated using optical laser fields yet are weakly coupled to the environment and hence have the long coherence lifetimes needed for optical memories and quantum computing. Recently, we performed proof-of-principle experiments in Pr doped $Y_2SiO_5$ (Pr:YSO) that demonstrated the potential advantage of Raman excited spin coherences for optical storage[4]. However, Pr:YSO has a weak optical oscillator strength ($\sim 10^{-7}$), as do many other spectral hole burning materials. This weak oscillator strength limits many potential applications because the ability to achieve efficient Raman excitation of a spin coherence depends on the product of the oscillator strength and the laser intensity. For example, for high temperature SHB applications, efficient Raman excitation becomes more difficult as the optical homogeneous width increases with temperature. Hence, for a fixed laser power, higher oscillator strength permits higher temperature operation. For quantum computing, the low optical transition rate limits the gate speed and the number of quantum logic operations that can be performed within the spin coherence lifetime[4].

For the current experiment we chose the N-V color centers in diamond because it has a large optical oscillator strength ($\sim 0.1$), relatively long spin coherence lifetimes (0.01-0.1 μsec), and has been previously demonstrated to exhibit Raman heterodyne signals[6]. The nitrogen-vacancy (NV) color center in diamond is created by radiation damage and annealing of Ib diamond, and is described elsewhere[7]. The sample we used is estimated to have about 30 ppm N-V color centers and has an peak optical density about 0.6 (25% transmission) for a 1 W/cm$^2$ probe intensity at 15 °K, on the zero-phonon line, near 637 nm optical wavelength. The optical density is much higher for weaker probe beam. The optical transition has an inhomogeneous broadening of 750 GHz, and a homogeneous width of 50 MHz. Previous work has shown that an applied magnetic field along (111) direction is required to observe Raman heterodyne signals[6], and therefore is needed for Raman excited spin coherences. Figure 1 shows an energy level diagram of the NV-center. The Raman transition frequency (~120 MHz) is determined by the spacing between the S=0 and S=-1 ground state spin sublevels. This spacing is controlled by the magnitude of the applied ~1 kGauss magnetic field. At this field strength, the S=0 and S=-1 ground sublevels for (111) oriented N-V centers are near an anti-crossing (Figure 1a). Under these conditions, a partial mixing of the spin sublevels makes it possible to enhance the Raman transition strength which otherwise would be small due to the small spin-orbit coupling[8].

The laser beams R1 and R2 in Figure 1b act as Raman pump beams to produce a two-photon ground state coherence via coherent population trapping[9]. In some of the experiments, a Raman enhanced non-degenerate four-wave mixing (NDFWM) technique is used to achieve a higher signal to noise ratio, in analogy to experimental techniques used previously to study Pr:YSO [4]. In this case, the laser beam P acts as a probe beam, which is diffracted from a phase grating produced by the two-photon coherence, and generates a diffracted beam D according to phase matching conditions **$K_D=K_{R2}-K_{R1}+K_P$**. To further enhance signal to noise, a heterodyne detection scheme is sometimes used to detect beam D. All the laser fields shown are derived from a single dye laser output using acousto-optic frequency shifters. This greatly relaxes dye laser frequency stability requirements since the resonant Raman interaction is insensitive to correlated laser jitter.

To match Figure 1, the Raman beams R1, R2, and probe beam P are downshifted from the original laser frequency 400, 280, and 420 MHz respectively. To generate lineshapes, the frequency of beam R2 was scanned around the 120 MHz Raman transition frequency with the frequencies of beams R1 and P held fixed. Compensating galvos allow changing the frequencies without perturbing the optical alignment. The intersection angle of the Raman beams was about 3.5° in the plane of the optical table. The direction of the probe beam P was optimized for diffraction efficiency but was 3.5° out of the plane of the optical table. An additional beam from the argon laser is also directed into the sample to serve as a repump. Without this repump beam, the NV center would exhibit persistent spectral hole-burning due to reorientation of the N-V center in the diamond lattice and no CW signal would be seen after a short time[10]. In contrast to our previous studies on Pr:YSO, the repump beam does not provide spectral selectivity for the four-wave mixing signal, therefore the optical transition is strongly inhomogeneously broadened. All laser beams were linearly polarized and focused into the crystal by a 150 mm focal length lens, producing a spot with a diameter about 100 μm. During the experiment, the sample was maintained at a temperature of 15 K inside the helium flow JANIS cryostat. The magnetic field was applied by a pair of Helmholtz coils that allow fine-tuning of the ground state energy level splitting near the anticrossing point. This magnetic field must be applied within 1° of the crystal (111)-direction and have a strength near 1 kGauss[6].

A representative NDFWM signal is shown in Figure 2 as a function of the frequency difference between Raman beams. The lineshape is taken at intensities substantially below saturation limits (see below). As shown, the Raman linewidth is about 5.5 MHz, which is comparable to the 5 MHz inhomogeneous width of the spin transition and much smaller than both the homogeneous width of the optical transition (~50 MHz) and laser jitter (~100 MHz). This sub-optical transition linewidth is taken as evidence of the Raman process. The maximal diffraction efficiency was found to be 0.5% for a very weak probe. This diffraction efficiency is reasonable given the probe detuning and the absorption due to high optical density of the sample.

To evaluate the relative matrix elements of our Raman system, we investigated the NDFWM signal amplitude as a function of Raman laser beam intensities. The results are shown in Figure 3. The saturation intensities were found to be 36 W/cm$^2$ and 56 W/cm$^2$ for the optical transitions R1 and R2 respectively. The fact that the observed saturation intensities for the Raman excited spin coherence are much larger than those predicted for a single isolated color center is due to the large inhomogeneous broadening of the optical transition. For the data of Figure 3, the intensity of the probe beam was 1.6 W/cm$^2$, which is substantially below its measured saturation intensity (48 W/cm$^2$). The intensity of the repump beam was around 10 W/cm$^2$.

Electromagnetically induced transparency[11] was also observed in the N-V diamond color center. To do this, the intensity of Raman beam (R2) was greatly reduced to 1 W/cm$^2$ and served as a probe, while the other Raman beam (R1) was increased to the maximal available intensity of 280 W/cm$^2$ and served as the coupling beam. The NDFWM probe beam (P) was blocked during the experiment. The experimental traces of the transmission of the probe as a function of Raman detuning are presented in Figure 4. The maximal value of transparency is 17% of the background absorption (optical density 0.3). The background absorption is reduced in the presence of the pump beam due to off-resonant pump-induced persistent spectral hole burning. We would like to point out that the observed value of transparency is large for such a strong inhomogeneous optical broadening. It is in good agreement with a simplified theoretical model which takes into account the inhomogeneous broadening and the fact that only 1 in 4 color centers has the correct orientation relative to the magnetic field. For applications, where single atoms will be excited, the theoretical model shows that resonant atoms with the correct orientation would exhibit close to 100% transparency under these experimental conditions (i. e. ~160 MHz Rabi frequency). Finally, the observed EIT linewidth of 8.5 MHz is substantially smaller than laser jitter and optical homogeneous linewidth is comparable to the inhomogeneous linewidth of the ground state spin transition.

In summary, we have observed electromagnetically induced transparency and nondegenerate four-wave mixing generation in an inhomogeneously broadened optically thick crystal of NV-diamond. The observed maximal value of EIT of 17% makes it possible to use this material in variety of EIT applications, such as non-linear image processing[12], optical data storage in solids, and solid state quantum computing. In particular, for quantum computing applications, we estimated that N-V diamond should be capable of more than 1000 logic gate operations ( ~160 MHz Rabi frequency) per spin decoherence time (~10-100 μs).

The authors are indebted to Dr. Steve Rand of University of Michigan for loan of the NV diamond crystal and to Dr. Neil Manson of Australian National University for valuable advice on experimental techniques. We also acknowledge discussions with S. Ezekiel of the Massachusetts Institute of Technology. This work was supported by ARO grant #DAAG55-98-1-0375, AFOSR grants #F49620-98-1-0313 and #F49620-99-1-0224.

**Figure captions.**

**Figure 1.** (a) Energy level splitting of N-V diamond as a function of magnetic field strength applyied in the (111) direction. (b) A Λ−shaped three-level system interacting with Raman beams and probe to generate a NDFWM signal at a magnetic field strength of about 1000G.

**Figure 2.** NDFWM signal efficiency at 15 K. Intensities of R1, R2, P, and Ar beams were 1.2, 1.6, 5.6 and 10 W/cm$^2$, respectively. Central difference frequency is 120 MHz.

**Figure 3.** Saturation curves for R1 and R2 beams (plots are vertically shifted for clarity). Intensity of the probe was 1.6 W/cm$^2$. The temperature of the sample was 15 K.

**Figure 4.** EIT amplitude relative to the peak probe absorption versus probe detuning in NV-diamond at 15 K for a coupling field intensity of 280 W/cm$^2$. The intensities of probe and repump were 1 and 10 W/cm$^2$ respectively. The sloping background is due to the frequency dependent efficiency of the acousto-optic shifter.

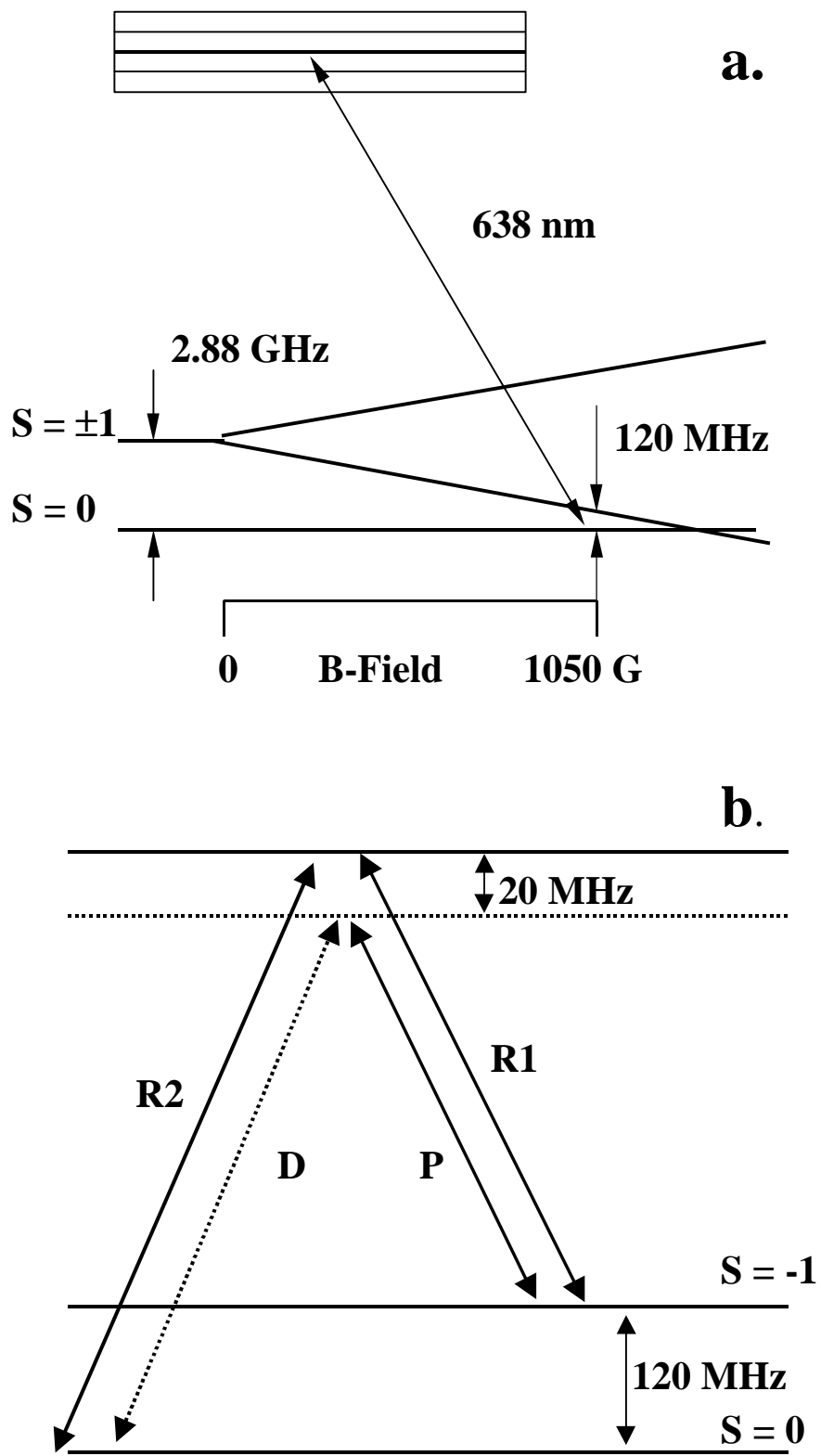

FIGURE 1

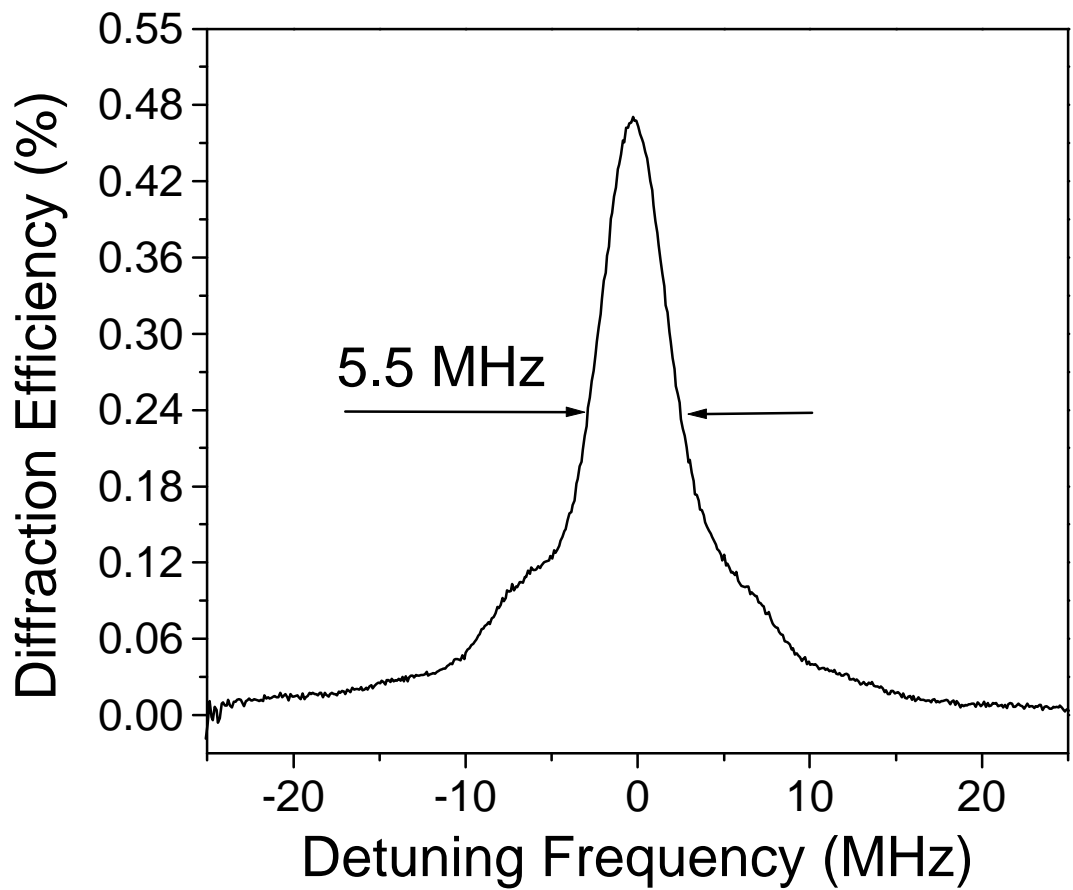

FIGURE 2

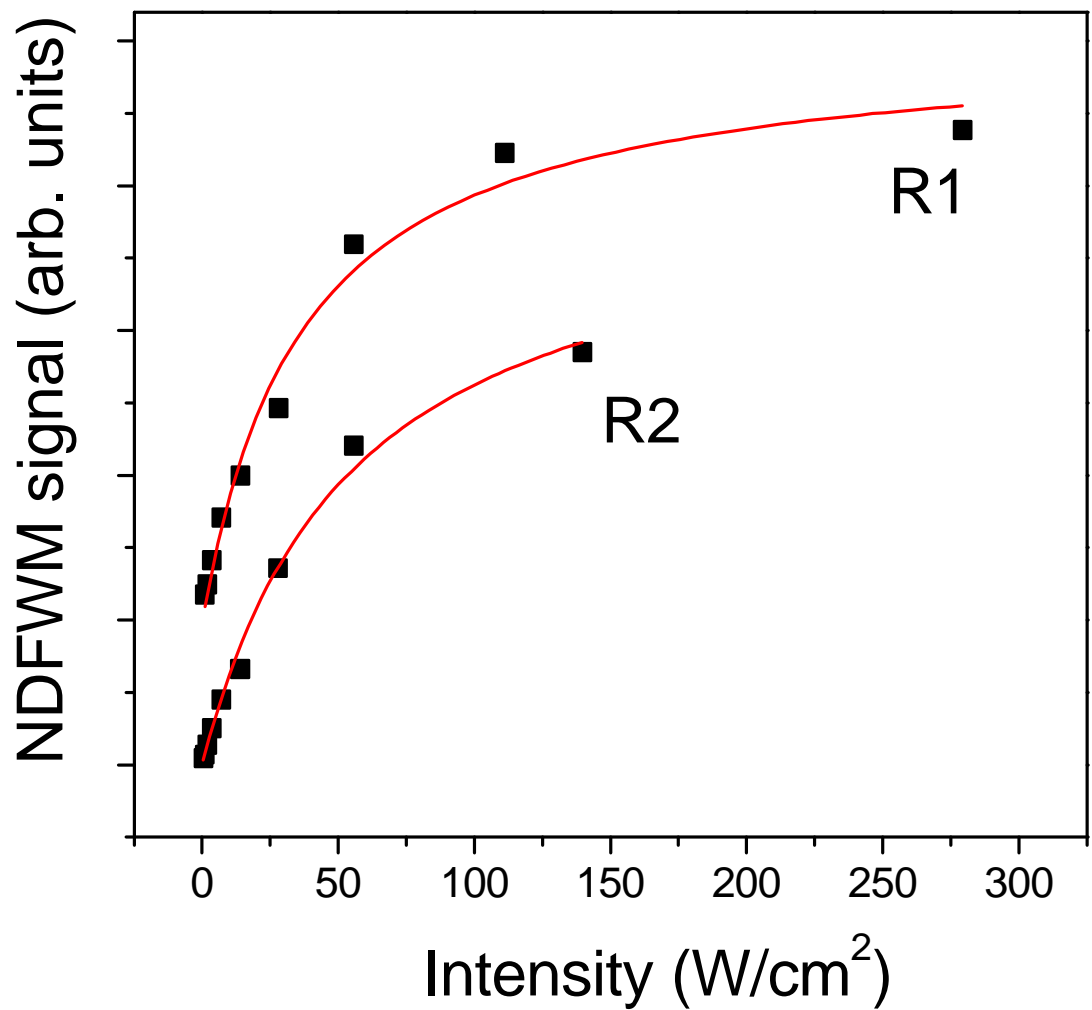

**FIGURE 3**

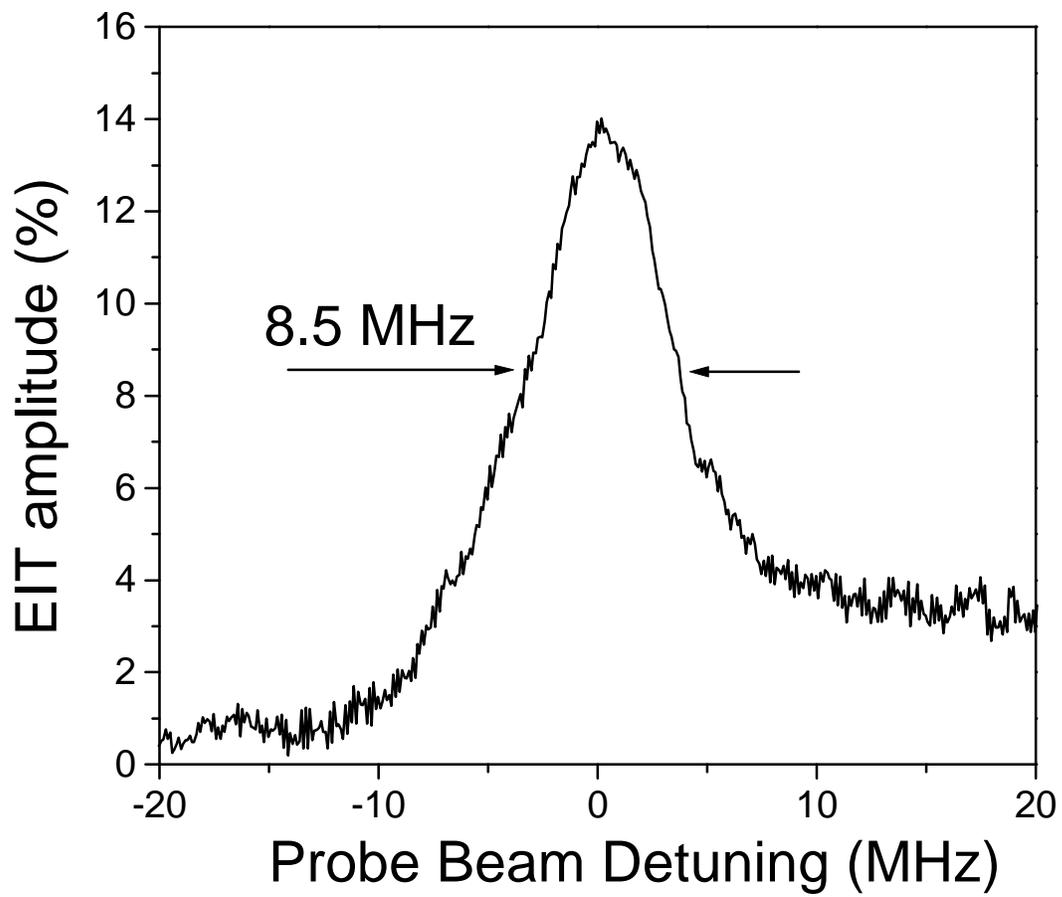

**FIGURE 4**